\documentstyle[12pt,prl,aps,multicol,psfig,epsf]{revtex}
\begin{document}
\title{
\hfill hep-ph/9706482\\
Neutrino Mass Constraints on R violation
and HERA anomaly}

\author{ Anjan S.~Joshipura,
         V.~Ravindran and  
         Sudhir K.~Vempati}
  \address{ Theoretical Physics Group, Physical Research Laboratory,
   Navarangpura, Ahmedabad, 380 009, India}
\def\ap#1#2#3{           {\it Ann. Phys. (NY) }{\bf #1} (19#2) #3}
\def\arnps#1#2#3{        {\it Ann. Rev. Nucl. Part. Sci. }{\bf #1} (19#2) #3}
\def\cnpp#1#2#3{        {\it Comm. Nucl. Part. Phys. }{\bf #1} (19#2) #3}
\def\apj#1#2#3{          {\it Astrophys. J. }{\bf #1} (19#2) #3}
\def\asr#1#2#3{          {\it Astrophys. Space Rev. }{\bf #1} (19#2) #3}
\def\ass#1#2#3{          {\it Astrophys. Space Sci. }{\bf #1} (19#2) #3}

\def\apjl#1#2#3{         {\it Astrophys. J. Lett. }{\bf #1} (19#2) #3}
\def\ass#1#2#3{          {\it Astrophys. Space Sci. }{\bf #1} (19#2) #3}
\def\jel#1#2#3{         {\it Journal Europhys. Lett. }{\bf #1} (19#2) #3}

\def\ib#1#2#3{           {\it ibid. }{\bf #1} (19#2) #3}
\def\nat#1#2#3{          {\it Nature }{\bf #1} (19#2) #3}
\def\nps#1#2#3{          {\it Nucl. Phys. B (Proc. Suppl.) }
                         {\bf #1} (19#2) #3} 
\def\np#1#2#3{           {\it Nucl. Phys. }{\bf #1} (19#2) #3}
\def\pl#1#2#3{           {\it Phys. Lett. }{\bf #1} (19#2) #3}
\def\pr#1#2#3{           {\it Phys. Rev. }{\bf #1} (19#2) #3}
\def\prep#1#2#3{         {\it Phys. Rep. }{\bf #1} (19#2) #3}
\def\prl#1#2#3{          {\it Phys. Rev. Lett. }{\bf #1} (19#2) #3}
\def\pw#1#2#3{          {\it Particle World }{\bf #1} (19#2) #3}
\def\ptp#1#2#3{          {\it Prog. Theor. Phys. }{\bf #1} (19#2) #3}
\def\jppnp#1#2#3{         {\it J. Prog. Part. Nucl. Phys. }{\bf #1} (19#2) #3}

\def\rpp#1#2#3{         {\it Rep. on Prog. in Phys. }{\bf #1} (19#2) #3}
\def\ptps#1#2#3{         {\it Prog. Theor. Phys. Suppl. }{\bf #1} (19#2) #3}
\def\rmp#1#2#3{          {\it Rev. Mod. Phys. }{\bf #1} (19#2) #3}
\def\zp#1#2#3{           {\it Zeit. fur Physik }{\bf #1} (19#2) #3}
\def\fp#1#2#3{           {\it Fortschr. Phys. }{\bf #1} (19#2) #3}
\def\Zp#1#2#3{           {\it Z. Physik }{\bf #1} (19#2) #3}
\def\Sci#1#2#3{          {\it Science }{\bf #1} (19#2) #3}
\def\n.c.#1#2#3{         {\it Nuovo Cim. }{\bf #1} (19#2) #3}
\def\r.n.c.#1#2#3{       {\it Riv. del Nuovo Cim. }{\bf #1} (19#2) #3}
\def\sjnp#1#2#3{         {\it Sov. J. Nucl. Phys. }{\bf #1} (19#2) #3}
\def\yf#1#2#3{           {\it Yad. Fiz. }{\bf #1} (19#2) #3}
\def\zetf#1#2#3{         {\it Z. Eksp. Teor. Fiz. }{\bf #1} (19#2) #3}
\def\zetfpr#1#2#3{         {\it Z. Eksp. Teor. Fiz. Pisma. Red. }{\bf #1} (19#2) #3}
\def\jetp#1#2#3{         {\it JETP }{\bf #1} (19#2) #3}
\def\mpl#1#2#3{          {\it Mod. Phys. Lett. }{\bf #1} (19#2) #3}
\def\ufn#1#2#3{          {\it Usp. Fiz. Naut. }{\bf #1} (19#2) #3}
\def\sp#1#2#3{           {\it Sov. Phys.-Usp.}{\bf #1} (19#2) #3}
\def\ppnp#1#2#3{           {\it Prog. Part. Nucl. Phys. }{\bf #1} (19#2) #3}
\def\cnpp#1#2#3{           {\it Comm. Nucl. Part. Phys. }{\bf #1} (19#2) #3}
\def\ijmp#1#2#3{           {\it Int. J. Mod. Phys. }{\bf #1} (19#2) #3}
\def\ic#1#2#3{           {\it Investigaci\'on y Ciencia }{\bf #1} (19#2) #3}
\def\tp{these proceedings}
\def\pc{private communication}
\def\ip{in preparation}
\relax

\newcommand{\GeV}{\,{\rm GeV}}
\newcommand{\MeV}{\,{\rm MeV}}
\newcommand{\keV}{\,{\rm keV}}
\newcommand{\eV}{\,{\rm eV}}
\newcommand{\Tr}{{\rm Tr}\!}
\renewcommand{\arraystretch}{1.2}
\newcommand{\beq}{\begin{equation}}
\newcommand{\eeq}{\end{equation}}
\newcommand{\beqa}{\begin{eqnarray}}
\newcommand{\eeqa}{\end{eqnarray}}
\newcommand{\ba}{\begin{array}}
\newcommand{\ea}{\end{array}}
\newcommand{\bmat}{\left(\ba}
\newcommand{\emat}{\ea\right)}
\newcommand{\refs}[1]{(\ref{#1})}
\newcommand{\ler}{\stackrel{\scriptstyle <}{\scriptstyle\sim}}
\newcommand{\ger}{\stackrel{\scriptstyle >}{\scriptstyle\sim}}
\newcommand{\lag}{\langle}
\newcommand{\rag}{\rangle}
\newcommand{\ns}{\normalsize}
\newcommand{\cm}{{\cal M}}
\newcommand{\gr}{m_{3/2}}
\newcommand{\p}{\partial}

\def\rp{ $R_P$} 
\def\321{$SU(3)\times SU(2)\times U(1)$}
\def\tl{{\tilde{l}}}
\def\tL{{\tilde{L}}}
\def\bd{{\overline{d}}}
\def\tL{{\tilde{L}}}
\def\a{\alpha}
\def\b{\beta}
\def\g{\gamma}
\def\c{\chi}
\def\d{\delta}
\def\D{\Delta}
\def\db{{\overline{\delta}}}
\def\Db{{\overline{\Delta}}}
\def\e{\epsilon}
\def\l{\lambda}
\def\n{\nu}
\def\m{\mu}
\def\nt{{\tilde{\nu}}}
\def\p{\phi}
\def\P{\Phi}
\def\x{\xi}
\def\r{\rho}
\def\s{\sigma}
\def\t{\tau}
\def\th{\theta}
\def\ne{\nu_e}
\def\nm{\nu_{\mu}}
\def\rp{$R_P$}
\def\mp{$M_P$}     
\renewcommand{\Huge}{\Large}
\renewcommand{\LARGE}{\Large}
\renewcommand{\Large}{\large}
\maketitle
\begin{center} Abstract \end{center}
\begin{abstract}
$R$ parity violating trilinear couplings $\lambda'_{1jk}$ of the
minimal supersymmetric standard model (MSSM)
are constrained from the limit on the electron neutrino mass.
Strong limits on these couplings follow from the earlier neglected
contribution due to sneutrino vacuum expectation values. The limits
on most of the $\lambda'_{1jk}$ couplings derived here are stronger than
the existing
ones for a wide range in parameters of MSSM. These limits strongly 
constrain the interpretation of recent 
HERA results in $e^+ p$ scattering in terms of production of
squarks through $R$ violating couplings. 
In particular, the interpretation in terms of
 $\tilde{t}_L$ production off  strange quark as suggested
recently is
not viable for  wide ranges  in MSSM
parameters.
\end{abstract}
\vskip 1.0truecm
The baryon and the lepton number symmetries are known to be violated when
the standard electroweak model is extended to include  supersymmetry.
These violations arise through  the following terms in the superpotential
of the minimal supersymmetric standard model (MSSM):
\begin{equation}
\label{wr}
W_{R} = -\tilde{\lambda}'_{ijk}~ L'_iQ'_{j} D'^{c}_{k}
-\tilde{\lambda}''_{ijk}~ U'^c_{i} D'^{c}_{j}
D'^c_{k}
-\tilde{\lambda}_{ijk}~ L'_{i} L'_{j}E'^c_{k}+\epsilon_i~ L'_iH_2~,
\end{equation}
where ${Q'}_{i} ,{D'}^{c}_{j}$ and ${L'}_{k}$ denote the
superfields corresponding to the weak eigenstates 
of quarks and leptons respectively.
The presence of the lepton number violating interactions leads to
interesting deviations from the standard model such as
$K^+\rightarrow\pi^+ \nu {\bar {\nu}}$, neutrino masses, violation of weak
universality
etc. The experimental limits on these have been used to constrain
\cite{sud8}
the relevant parameters in eq.(1). A particularly stringent
constraint
follows from the limit on the electron neutrino mass.
The existing analysis \cite{sud10,sud11} of these constraints has
neglected an important
and dominant contribution to the neutrino masses which would occur 
when any of the lepton number violating term in eq.(1)
is present. Incorporation of this additional contribution can change 
\cite{carlos} the
already obtained \cite{sud10,sud11}
limits significantly. The aim of this paper is to systematically
derive these constraints and discuss its implications.

A careful analysis of the constraints on $\tilde{\lambda}'$ couplings is
of
particular significance in view of the recent anomaly observed by the
HERA experiments H1 \cite{sud1}  and ZEUS \cite{sud2}. They
 have reported excess events in deep inelastic $e^+p$ scattering at 
$Q^{2} > 1.5 \times 10^{4}~ \GeV^{2}$.  This excess cannot be reconciled 
with the expectations based on the QCD improved parton model and point 
to the existence of physics beyond standard model.  The reported excess 
could be attributed to the presence of non-standard contact 
interactions\cite{sud3}, to the resonant production of a leptoquark 
state \cite{sud4} or of a squark through its R parity violating couplings
$\lambda'$ \cite{sud5,sridhar}. 

The possible resonance seen at HERA can be identified either with a
$\tilde{c}_L$ or a $\tilde{t}_L$ squark of the MSSM. Three plausible
production mechanisms have been proposed \cite{sud5,sridhar} and
discussed:  $e^+_Rd_R\rightarrow \tilde{c}_L,e^+_Rd_R\rightarrow
\tilde{t}_L,e^+_Rs_R\rightarrow \tilde{t}_L$. All these mechanisms
(particularly the last one) require
significantly large values for $\lambda'_{1jk}$ which are
constrained among other things by the electron neutrino
mass.  

The neutrino masses arise through two different sources\cite{sud9} in the 
supersymmetric
standard model. The lepton number violating trilinear terms in eq.(1)
constitute the first source. They generate majorana masses at the
one-loop level 
\cite{sud9,1loop}.
 The lepton number violating terms in superpotential are invariably
accompanied 
by similar terms in the scalar potential providing 
another source of neutrino masses. Specifically, the following 
 terms are allowed by $SU(3) \times SU(2) \times U(1)$ invariance in V:
\begin{equation}
V_{soft} = - B_{\tilde{\nu}_{i}}~ \tilde{\nu}_{i} H_{2}^{0}  
+  m^{2} _{\nu_{i} H_{1}}~  \tilde{\nu}^{\star}_{i} H_{1}^{0} + c.c +\cdots~.
\end{equation}
These terms are present at the GUT scale if original 
superpotential contains bilinear R violating terms. 
But terms in eq.(2) would be generated \cite{carlos,sud15}
at the weak scale even when the violation of lepton number is only through the 
trilinear terms as in eq.(1).
The major consequence of these terms is the generation of sneutrino VEV which
causes the mixing of neutrinos with gauginos leading to the neutrino masses
\cite{carlos,sud15,sud13,sud14,sud16}.

The bounds on $ \tilde{\lambda'}_{1ij}$ quoted \cite{sud8,sud5} in the literature 
are derived \cite{sud10,sud11}
using  the loop induced contribution to the neutrino masses. 
Recent analysis of neutrino masses 
\cite{carlos,sud15,sud13,sud14,sud16}
has revealed that the 
neutrino gaugino mixing tends to generate a 
larger neutrino mass compared to the loop induced contribution. 
This contribution can thus lead to 
stronger bounds on the trilinear couplings. 
We derive these bounds. It turns out that for a range of the MSSM
parameters the bounds on $\lambda_{1ij}'$
following from the neutrino mass can be  stronger than the existing
bounds based on other processes such as $K^+\rightarrow \pi^+
\nu\bar{\nu}$ \cite{sud17}, neutrinoless double beta decay \cite{db},
$Z\rightarrow e^+e^-$ \cite{sridhar1} etc.
As a result, a sizable part of the parameter space otherwise
allowed for the interpretation of HERA events is curtailed by the neutrino
mass bounds.

We work with R violating version of the MSSM with a standard  set of soft 
supersymmetry breaking terms
specified at a high scale near $ M_{GUT} = 3 \times 10^{16}~ GeV$. 
Since our aim is to derive bounds  on $\tilde{\lambda'}$ couplings, we
concentrate only on that term in eq.(1) and set other couplings to zero. 
It is more convenient to 
express this equation in terms of the mass eigenstates $Q_{i},D^{c}_{j}$
and $L_{k}$ of quarks and charged 
leptons  respectively\cite{sud17}.
\begin{equation}
W_R = - \lambda'_{ijk}~ E_{i} U_{j} D^{c}_{k} - \lambda^{\nu}_{ijk}~\nu_{i}
D_{j} D_{k}^{c} \;\;,
\end{equation}
where $\lambda'_{ijk}$ are related \cite{sud17} to 
$\tilde{\lambda}_{ijk}$ of eq.(1) and 
\begin{equation}
\lambda^{\nu}_{ijk}\equiv V^{KM}_{lj}~ \lambda'_{ilk}~.
\end{equation}
The lepton number violating part of the scalar potential at the weak
scale is given by 
\begin{equation}
V_{soft} = A^{\nu}_{ij} \lambda^{\nu}_{ij}~ D_{i} D_{j}^{c} \tilde{\nu} -
B_{\nu}~ \tilde{\nu} H_{2} + m^{2}_{\nu H_{1}}~ \tilde{\nu}^{\star}  H_{1}
+ c.c + \cdots ~,
\end{equation}
where we have explicitly written only those terms which are responsible
for generating the electron neutrino ($\nu_{1} \equiv \nu$) mass and 
$\lambda^{\nu}_{ij}\equiv \lambda^{\nu}_{1ij}$.
The terms in eq.(5) generate a non-zero VEV$\langle\tilde{\nu}\rangle$ and 
give rise to the following neutrino mass 
\cite{sud9,sud15,sud13,sud14,sud16,sud12}:
\begin{equation}
(m_{\nu_{e}})_{tree}\simeq \frac{\mu ~(c g^{2} + {g'}^2)~ \langle\tilde{\nu}\rangle^{2}} {4 ~( - c~ \mu M + M_{w}^{2}~ \cos\beta ~\sin\beta ~(c + \tan^2\theta_{w}))}~,
\end{equation} 
where $c\simeq 5 {g'}^{2}/3 g^2$, $g^2$ and
${g'}^{2}$
are gauge couplings and $\tan\beta = \langle H_{2}^{0} \rangle/
\langle H_1^0 \rangle  = 
v_2/v_1$. 
The $\langle \tilde{\nu} \rangle$ follows from the
minimization of the scalar potential which contains terms of
eq.(5) in addition to the standard soft terms of MSSM. Explicitly,
\begin{equation}
\langle \tilde{\nu} \rangle \sim {\displaystyle{{B_\nu v_{2} - m_{\nu H_{1}}^2 v_{1}}} \over 
{m_{\tilde{\nu}}^{2} + \frac{1}{2} M_{z}^{2} \cos 2 \beta}}~,
\end{equation}
$m_{\tilde{\nu}}^{2}$ is the soft SUSY breaking sneutrino mass.

All  parameters in eq.(6) are specified at a low scale $Q_{0}$ . 
We have chosen $Q_{0}$ to be $M_{Z}$ \cite{sud19} . The parameters 
$B_{\nu},m_{\nu H_{1}}^{2}$ are assumed to be zero at $M_{GUT}$ .
 Their values at $Q_{0}$ follow from the following 
renormalisation group(RG) equations \cite{sud20} :
\begin{eqnarray}
\frac{dB_{\nu}}{dt}& =& -~ \frac{3}{2} B_{\nu}~\left(Y_{t}^{U} - \tilde{\alpha}
_{2} - \frac{1}{5} \tilde{\alpha}_{1}\right) - \frac{3 \mu}{16 \pi^{2}} 
{\lambda }_{kk}^{\nu} h^{D}_{k} \left({A}^{\nu}_{kk} + \frac{1}{2} B_{\mu} \right)~, \\
\frac{d m_{\nu H_{1}}^{2}}{dt}& = &-~ \frac{1}{2} m_{\nu H_{1}}^{2} \left(3 Y_{k}^{D} + 
Y_{k}^{E} \right)  
-\frac{3}{32 \pi^{2}} {\lambda }^{\nu}_{kk}
h^{D}_{k} \left( m_{H_{1}}^{2} + m_{\tilde{\nu}}^{2} \right. \nonumber \\
& & \left. +2~ A_{kk}^{\nu} A_{k}^{D} + 
2~  m_{k}^{Q{2}} +~ 2 ~m_{k}^{D^{c}2 } \right)~.
\end{eqnarray}
In the above equations, $ t\equiv 2~ ln(M_{GUT}/Q_0) $;
$h_{k}^{D},h_{k}^{U},h_{k}^{E} $ are Yukawa couplings 
corresponding to the down,up quarks and the charged leptons respectively ; 
$Y_{k}^{D,U,E}\equiv (h_{k}^{D,U,E})^2/16
\pi^2$ ; $\tilde{\alpha}_{i} \equiv \alpha _ {i} /4 \pi$
.  The sum over generation indices $k,l$ is implied above. The parameters
 $\lambda_{kk}^{\nu}$  and $A_{kk}^{\nu}$ satisfy the following RG equations:
\begin{eqnarray}
\frac{d \lambda^\nu_{kk}}{dt}& = &  \lambda^\nu_{kk}\left(- 3 Y^D_k - {1 \over 2} Y^U_k  + {7 \over 30} \tilde \alpha_1 + {3 \over 2} \tilde \alpha_2 + {8 \over 3}
\tilde \alpha_3 \right), \nonumber\\
\frac{d A_{kk}^{\nu}}{dt}& =& {3 \over 2} A_{kk}^\nu Y_k^D -
{9 \over 2} A_k^D Y_k^D - {1 \over 2 } A_k^U Y_k^U - \frac{16}{3} \tilde \alpha_{3} M_{3}  - 3 \tilde \alpha_{2}
M_{2} - \frac{7}{15}\tilde  \alpha_{1} M_{1}~. 
\end{eqnarray}

We have kept only leading order terms in $\lambda_{kk}^{\nu}$ 
in writing eqs.(8 -10). Rest of the parameters appearing in eqs. (8 -10)
 satisfy the standard RG equations to this order in $\lambda_{ij}^{\nu}$.
Note that the second term in eqs.(8-9) generate non-zero  $B_{\nu}$
and $m_{\nu H_{1}}^{2}$ at $Q_{0}$.
The neutrino mass $(m_{\nu_{e}})_{tree}$ following from eqs.(6 -10) 
thus involves the combination $(\lambda_{kk}^{\nu} h_{k}^{D})^{2}$.

The trilinear interactions in eq.(3) lead  to the following
$m_{\nu_{e}}$ at the one-loop level \cite{sud9,1loop}:
\begin{equation}
(m_{\nu_{e}})_{loop} \simeq -~ \frac{(\lambda_{kk}^{\nu})^{2}}{16 \pi^{2}}
m_{k}^{D} \sin\phi_{k} \cos\phi_{k}~ ln \frac{M_{2 k}^{2}}{M_{1 k}^{2}}~,
\end{equation}
where we have implicitly assumed that only one $\lambda'_{1jk}$ is
non-zero at a time. $\phi_{k}$ and $M_{2,1 ~ k}^{2}$ respectively denote 
the mixing among
 squarks $\tilde{d}_{k},\tilde{d}^{c}_{k}$ and their masses. 
The mixing $\phi_{k}$  is proportional to $m_{k}^{D}$  and hence 
the $(m_{\nu_{e}})_{loop}$ also scales as 
$(\lambda_{kk}^{\nu} h_{k}^{D})^{2}$. $m_{\nu_e}$ therefore provides a
bound on $\lambda_{kk}^{\nu}$ which can be converted to a bound on
$\lambda'_{1lk}$ using ~eq.(4). The resulting bounds
become 
stronger with increase in $\tan\beta$
due to the fact that $B_{\nu},m_{\nu H_{1}}^{2} $ 
involve $h_{k}^{D} = m_k^D/v \cos\beta$ .  
For the same reason, bounds  display strong hierarchy, typically
\begin{equation}
\frac{(\lambda_{kk}^{\nu})_{max}}{(\lambda^{\nu}_{ll})_{max}}\sim\frac{m_{l}^{D}}
{m_{k}^{D}}~.
\end{equation}
It also follows from eq.(4) that
\begin{equation}
\frac{(\lambda'_{1kk})_{max}}{(\lambda'_{1lk})_{max}}\sim\frac{V_{lk}^{KM}}
{V_{kk}^{KM}}~.
\end{equation}
\noindent
It is seen from last two equations that the
$\lambda'_{133},(\lambda'_{131})$
is constrained most (least) by $m_{\nu_{e}}$.

The Kobayashi Maskawa mixing neglected in the earlier analysis plays 
 quite an important role here which needs to be emphasized. The earlier
analysis
could not constrain $\lambda'_{1jk}$ with $j\neq k$ due to the fact that
single such $\lambda'_{1jk}$ leads to zero
$m_{\nu_e}$ when $V_{KM}$ is neglected. This is no longer true in the
presence of $V_{KM}$ allowing us to
derive significant bounds on couplings $\lambda'_{1jk}\;\; (j\neq k)$.

We have numerically solved eqs.(8-10) and the
standard equations of other 
parameters appearing
in them. They lead to  $B_{\nu}$ and 
$m_{\nu H_{1}}^{2} $  at $t_{0} = 2~ln~\frac{M_{GUT}}{M_Z} $ which are 
 specified in terms of the standard MSSM parameters which 
we choose as gaugino (gravitino) mass, $M_2\;(m)$, $\tan\beta$ and
universal
 trilinear strength $A$. $B_{\nu}$ and $m_{\nu H_{1}}^{2}$ determine 
$(m_{\nu})_{tree}$ in terms of these parameters and 
$\lambda_{kk}^{\nu}(t_{0})$. The 1-loop contribution also gets
fixed in terms of these parameters since $\phi_{k}$ and
$M_{2,1~k}^{2}$ appearing
in eq.(11) are determined using the standard $2 \times 2$  mixing matrix for 
$\tilde{d}_{k},\tilde{d}^{c}_{k}$  system.  

The  bounds  on different couplings are displayed in Figs. 1 and 2. 
Apart from being dependent on SUSY parameters, these bounds are quite
sensitive to the chosen sign of the $\mu$ parameter. This is due to the
fact that for one (namely
negative) sign of $\mu$, two terms appearing in the sneutrino VEV, eq.(7)
cancel while they add for positive sign. The suppression in sneutrino VEV
occurring in the first case weakens the bound on $\lambda'$. Such
suppression has been previously noted in the case of bilinear $R$ violation
\cite{carlos,sud13,sud14,sud16}.
Fig. 1 displays bounds on couplings $\lambda'_{111}$ and $\lambda'_{121}$
obtained by demanding $m_{\nu_e}\equiv
(m_{\nu_e})_{tree}+(m_{\nu_e})_{loop}\leq 5 \eV$. Curves for three
representative values of $\tan\beta$ and universal gravitino mass
$m$ are shown. For comparison, we also display in the same figure
the existing bounds on these couplings. The most
stringent bound on $\lambda'_{111}$ is derived from neutrinoless
double beta decay
and on  $\lambda'_{121}$ from the process $K^+\rightarrow \pi^+\nu \bar{\nu}$.
These are shown as function of $M_2$ in the same figure using MSSM
expressions for the relevant squark masses. It is seen that the bounds
derived here are comparable or better (in case of  larger
values of $M_2$ ) than the
already existing ones.

Fig 2a ($\mu>0$) and 2b ($\mu<0$) show the bounds on $\lambda'_{133}$
and $\lambda'_{132}$. The $\lambda'_{133}$ is constrained most and 
is required to be as low as $ 3\times 10^{-6}$ even for $\tan\beta=5$. For
comparison we also show the limit on $\lambda'_{133}$ which would follow
if sneutrino VEV is completely neglected as in previous works
\cite{sud10,sud11}. It is clear
that inclusion of this VEV drastically alters the bound. We also display
limit on $\lambda'_{132}$ which was poorly constrained in the previous
analysis \cite{sridhar1} and which is relevant for the interpretation of
the HERA events. 
\noindent
The
limits displayed in fig 2a are at least an order of magnitude stronger
than the value needed to reconcile HERA anomalies. 
The resulting bounds on other
couplings $\lambda'_{1jk}$ not explicitly displayed in above figures can
be read off from eq.(13). These constraints are listed in a table for
two representative values $m=M_2=50 \GeV$ and $m= 50\GeV, M_2=75 \GeV$
and $\tan\beta=10$.
These respectively correspond in MSSM to $m_{\tilde{d}_R}\sim 163,240 
\GeV$. It is seen that the neutrino mass considerably improves
on the existing constraints.

\begin{center}
\begin{tabular} {|l|l|l|l|}
\hline
& &$~~m=50~GeV~~$ &$~~m=50~GeV~~$ \\
$~~\lambda'_{ijk}~~$& $~~$Previous limits$~~$ &$~~M_2=50~GeV~~$ &$~~M_2=75~GeV~~ $ \\
\hline \hline
\quad 111 &\quad 0.002\cite{db}    &\quad 0.0012           &\quad 0.0052    \\
\quad 121 &\quad 0.024\cite{sud17} &\quad 0.0056           &\quad 0.023     \\
\quad 131 &\quad 0.52\cite{barger} &\quad 0.015            &\quad 0.065     \\
\quad 112 &\quad 0.024\cite{sud17} &\quad 0.0003           &\quad 0.0012    \\
\quad 122 &\quad 0.024\cite{sud17} &\quad 6.3 $~10^{-5}$  &\quad 2.6 $~10^{-4}$  \\
\quad 132 &\quad 0.56\cite{sridhar}&\quad 0.0015           &\quad 0.0065       \\
\quad 113 &\quad 0.024\cite{sud17} &\quad 2.3 $~10^{-5}$   &\quad 9.8 $~10^{-5}$\\
\quad 123 &\quad 0.024\cite{sud17} &\quad 4.6 $~10^{-5}$   &\quad 1.3 $~10^{-4}$ \\
\quad 133 &\quad 0.002\cite{sud10} &\quad 1.9 $~10^{-6}$   &\quad 7.9 $~10^{-5}$ \\
\hline
\end{tabular} \\
\end{center}
{\bf Table 1.}{\it { Limits on single
$\lambda'_{ijk}$ following from the electron neutrino mass for 
$A=0$,\\ $\tan\beta=10$ and $\mu>0$. The limits become stronger  for
larger $\tan\beta$ and weaker for\\ negative $\mu$. The existing
limits mentioned in the table are for the relevant squark \\mass $\sim 200
\GeV$ and gluino mass $M_3=200~GeV$ in case of $\lambda'_{111}$ }} \\

\vskip .5cm
We now briefly discuss the implication of the above bounds on the
interpretation
of HERA events deferring details to a latter work. The H1 and
ZEUS data allow  interpretation as resonance production of squarks 
provided \cite{sud5,sridhar}
\begin{eqnarray}
(\lambda'_{121},\lambda'_{131})\sim \frac{.04}{B^{1/2} }& \;\;(e^+_R
d\rightarrow \tilde{c}_L, \tilde{t}_L)~, \nonumber \\
\lambda'_{132}\sim \frac{0.3} {B^{1/2}\cos\phi}& \;\;(e^+_R
s\rightarrow  \tilde{t}_L)
\end{eqnarray}
where $B$ refers to the branching ratio for the squark decay to $q e^+$
and $\phi$ is the $\tilde{t}_L-\tilde{t}_R$ mixing angle.
Limits on couplings displayed in Figs 1 and 2 are indeed stronger than
above
(even
for $B\sim 1$) for a large range of MSSM parameters. In
particular, the interpretation in terms of the $\tilde{t}_L$ production 
off strange quark at HERA is  severely constrained
from $m_{\nu_e}$. This possibility was studied in detail in \cite{sridhar}
which included constraints on $B$ from CDF, LEP constraints on
$\delta\rho$ and restrictions on $\lambda'_{132}$ coming from
$Z$ partial widths. The last one gives fairly weak constraint
on $\lambda'_{132}$ compared to the one coming from $m_{\nu_e}$. This
constraint therefore places strong restrictions on the available
parameter space. These  are displayed in Fig. 3 where we 
show the
region in $m-M_2$ plane in which neutrino mass bound is satisfied and 
$\lambda'_{132}>0.3$. 
We have also displayed the restrictions coming
from the experimental bound of 85 GeV on the chargino masses at LEP. The
combined allowed region becomes smaller as $\tan\beta$
increases and eventually beyond $\tan\beta=10 $ it gets
pushed
to $m\geq 240 \GeV$ \cite{fn}. 
This  will be limited further when one includes other
constraints such as $\delta\rho$ and CDF constraints on $B$ relevant, when
the squark mass is below 210 GeV. The corresponding restrictions are
much stronger
when $\mu>0$. For example, we find in this case that neutrino mass bound
is
inconsistent with $\lambda '_{131}\geq 0.3$ for $m,M_2\leq 600
\GeV$ \cite{fn} and $\tan\beta\geq 1$ if $A=0$. 
Somewhat milder but still significant restrictions are also placed by
$m_{\nu_e}$ on the alternative explanation in terms of production
of a charm squark off $d$. Fig. 3 also shows curves corresponding
to $\lambda'_{121}=0.04$ for $\tan\beta=40$. The allowed region expands
for lower $\tan\beta$. Here again, other constraints considered in
\cite{sud5} would further limit the available space. 
In contrast to these two case, the process $e^+ d\rightarrow \tilde 
{t}_L$ is not significantly constrained by $m_{\nu_e}$ as bounds on the relevant
coupling $\lambda'_{131}$ are weak due to appearance of Cabibbo angles. 
It is clear  that  
while the neutrino mass bounds may not rule out the SUSY interpretation
of HERA events they provide by far the strongest constraints 
 which should be systematically included in any analysis
trying to understand HERA events as the production of a squark.

All our considerations are based on specific set of soft terms
which are included  in MSSM. Many of the discussions and in particular
the generation of sneutrino VEV in the presence of a single trilinear
coupling would be true in a more general set up, {\em e.g.} in
case when soft terms arise from the gauge mediated supersymmetry breaking.
The main point
following from our analysis is that the electron neutrino mass provides
much stronger constraints on $R$ violation and associated phenomenology
than has been hitherto realized.  

\newpage
\begin{figure}[h]
\epsfxsize 15 cm
\epsfysize 15 cm
\epsfbox[25 151 585 704]{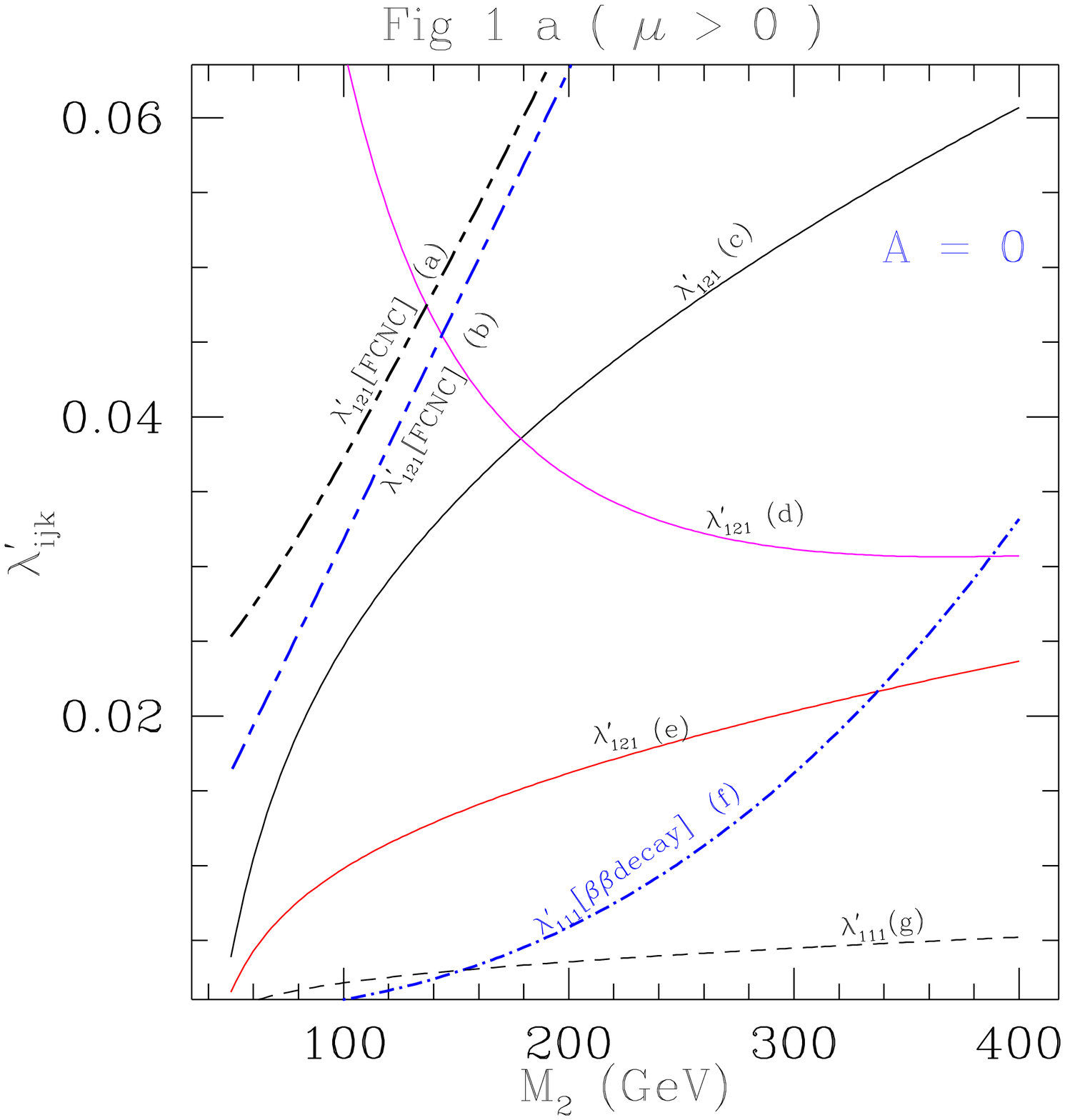}
\label{fig1a}
\end{figure}
\noindent
{\bf Figure 1a}.~\sl{ FCNC constraints [17] on $\lambda^{\prime}_{121}$ for a)~$m = 200$ GeV and 
b)~ $m = 50$ GeV for $tan\beta = 40$.
Neutrino mass constraints on $\lambda^{\prime}_{121}$ for 
c)~$m = 50$ GeV, $tan\beta = 15$; d)~$m = 200$ GeV ,$tan\beta = 40$ and	 
e)~$m = 50$GeV, $tan\beta = 40$.	 
f)~ $\lambda^{\prime}_{111}$ from neutrino less $\beta\beta$ decay 
g)~$\lambda^{\prime}_{111}$ from neutrino mass constraints for $m = 50$ GeV 
and $tan\beta = 40$.}
\newpage
\begin{figure}[h]
\epsfxsize 15 cm
\epsfysize 15 cm
\epsfbox[25 151 585 704]{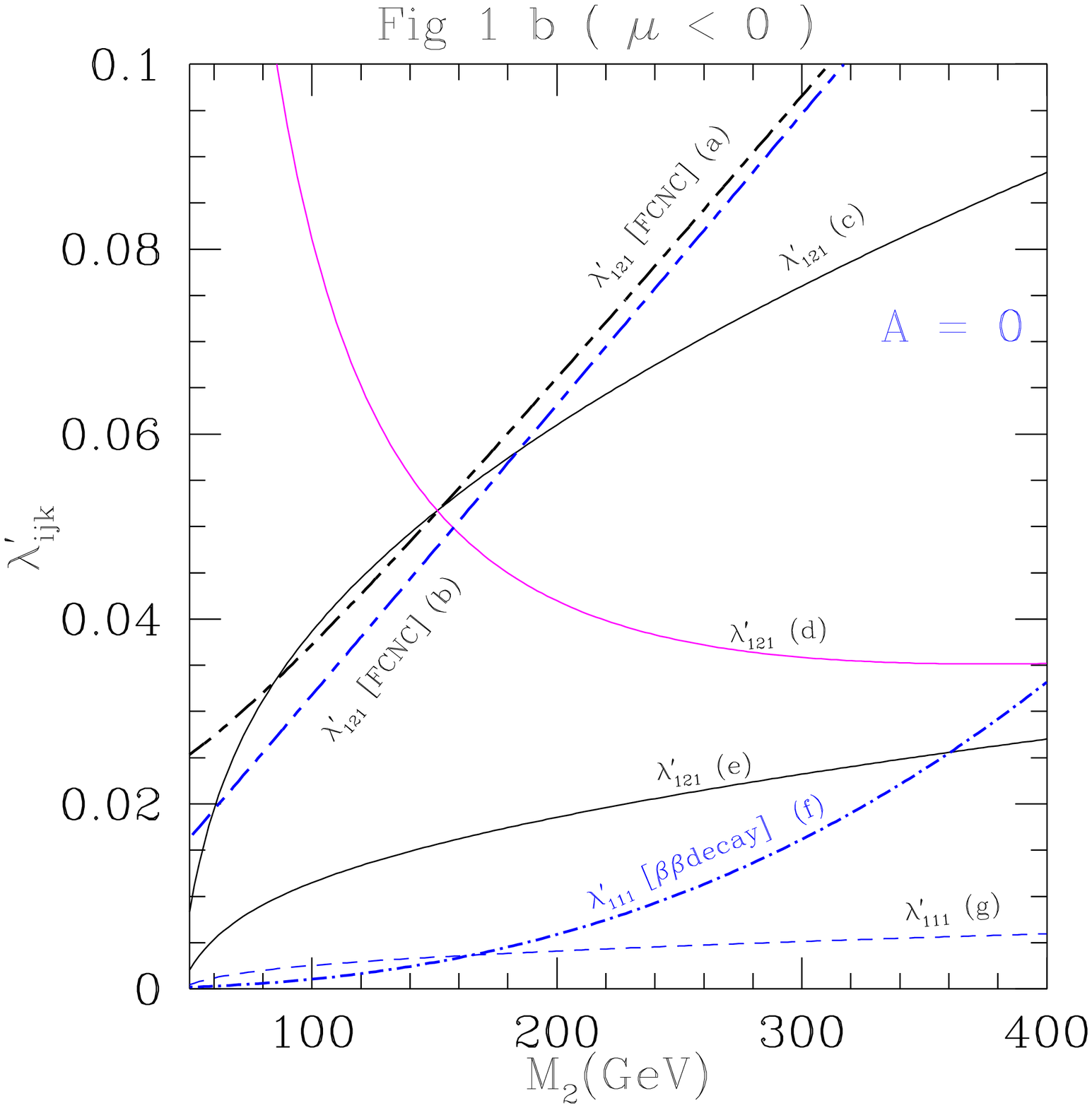}
\label{fig1b}
\end{figure}
\noindent
{\bf Figure 1b}.~\sl{ Same as of Figure {\bf 1a)} but with $\mu <0$ }
\newpage
\begin{figure}[h]
\epsfxsize 15 cm
\epsfysize 15 cm
\epsfbox[25 151 585 704]{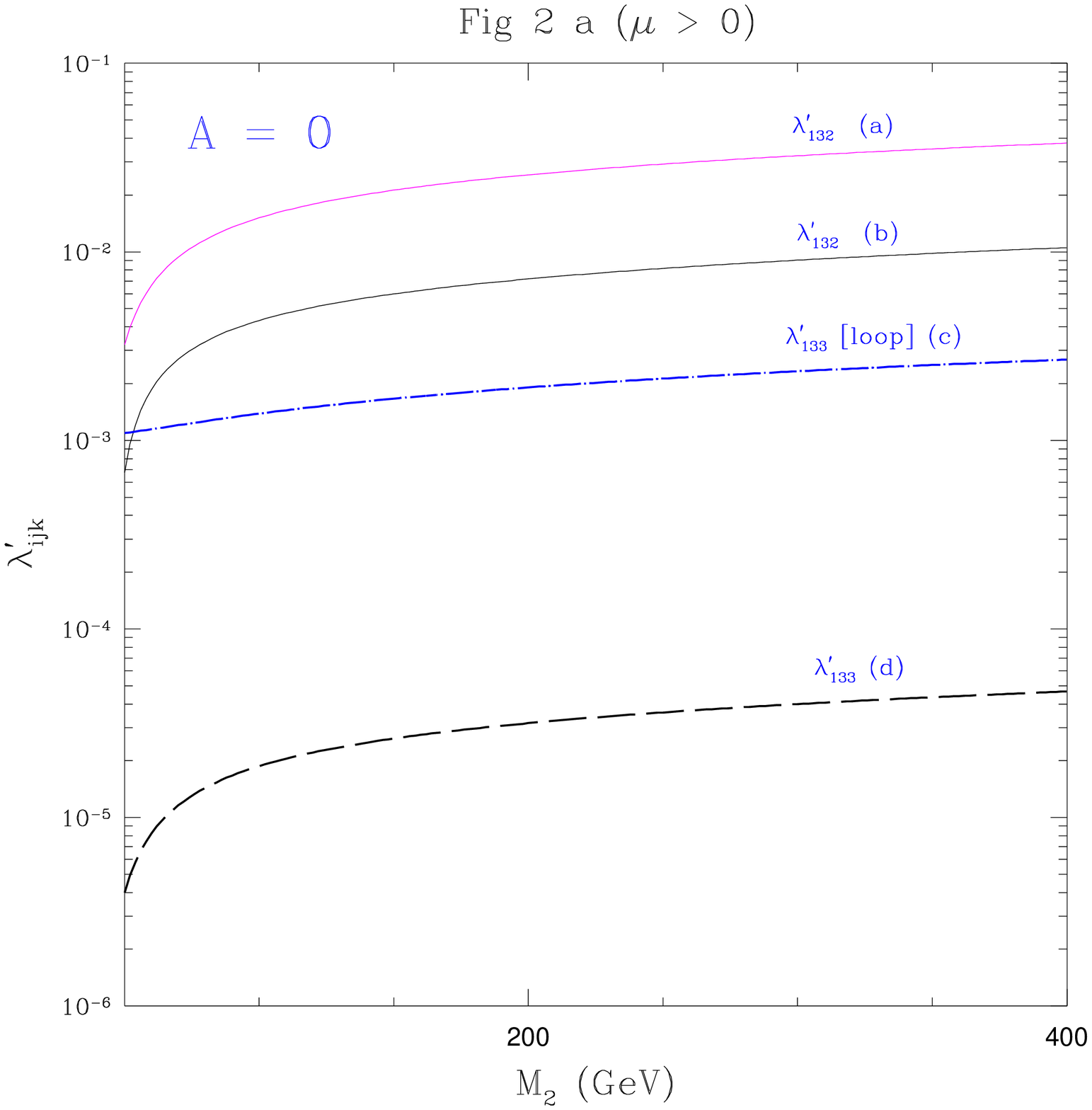}
\label{fig2a}
\end{figure}
\noindent
{\bf Figure 2a}.~{\sl 
Neutrino mass constraints on $\lambda^{\prime}_{132}$ for a)~$tan\beta = 5$
and b)~$tan\beta = 25$; on $\lambda^{\prime}_{133}$ for $tan\beta = 5$ 
c)~considering only loop contributions and d)~loop as well as sneutrino VEV 
contributions are shown for $m=50~GeV$. 
}
\begin{figure}[h]
\epsfxsize 15 cm
\epsfysize 15 cm
\epsfbox[25 151 585 704]{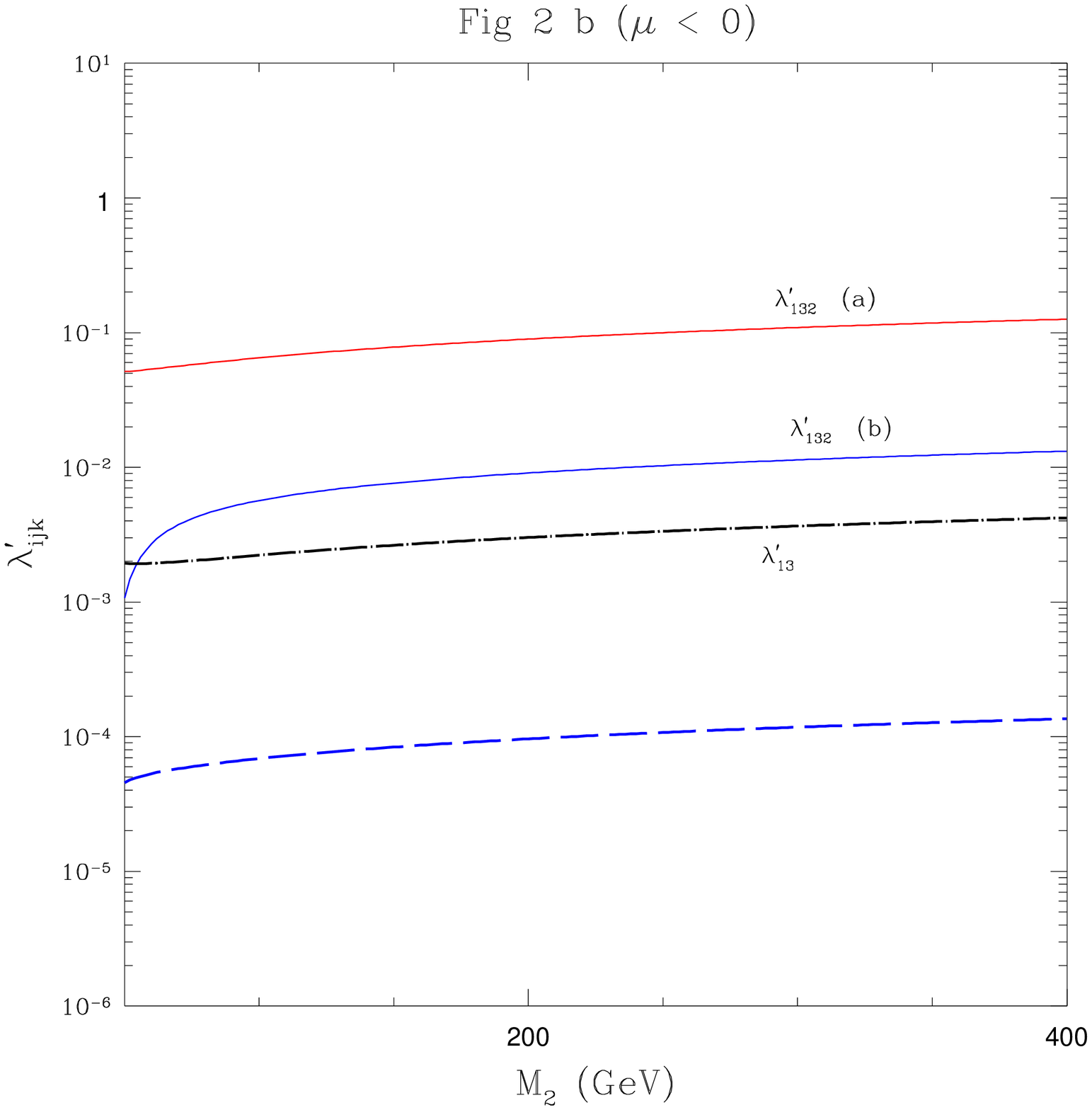}
\vspace*{.5cm}
\label{fig2b}
\end{figure}
\noindent
{\bf Figure 2b}.~{\sl Same as Figure {\bf 2a)} but with $\mu < 0$}. 
\newpage
\begin{center}
\begin{figure}[h]
\hspace*{.5 cm}
\epsfxsize 15 cm
\epsfysize 15 cm
\epsfbox[25 151 585 704]{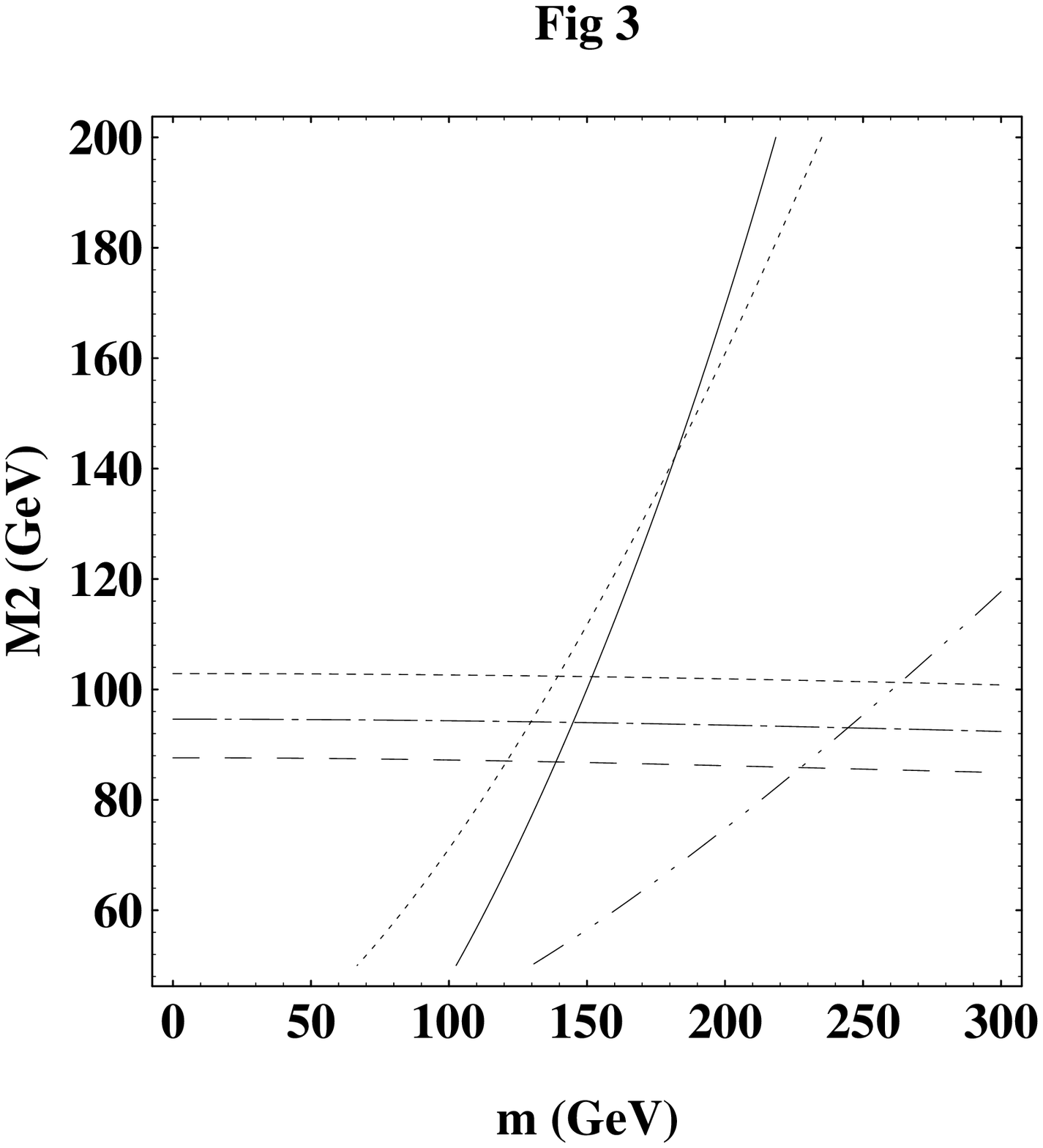}
\vspace*{.6cm}
\label{fig3}
\end{figure}
\end{center}
\noindent
{\bf Figure 3}.~{\sl 
The horizontal lines correspond to contours of $m_{\chi} = 85$GeV for 
$tan\beta = 40$ and $\mu>0$(small dashed), 
$tan\beta = 10$ and $\mu<0$ (dash-dot), $tan\beta = 5$ and $\mu<0$(dashed).
Contours of $\lambda_{132}^{\prime} = 0.3$ for $tan\beta = 5$
and $\mu<0$ (continuous line), 
$tan\beta = 10$ and $\mu<0$ (dash-double dot) are also shown. 
The dotted line corresponds to $\lambda_{121}^{\prime} = 0.04$ 
for $tan\beta = 40$ and $\mu>0$.  The region to the left (below) of the
contour (the horizontal line) is excluded.
}

\begin{thebibliography}{99}
\bibitem{sud19} Note that change in $Q_0$ can alter some of the bounds
obtained by minimization of the tree level potential and more detailed
treatment should include 1-loop corrected potential, see G. Gamberini, G.
Ridolfi and F. Zwirner, \np{B331}{90}{331}.
\bibitem{sud20} The relevant RG equations follow from the general expressions
given by N. K. Falck, \Zp{C}{86}{247}. Some of these have also
been given in \cite{carlos}.
\bibitem{fn} Note that values of $m, M_2>O(200 \GeV)$ will not be relevant
for
explaining HERA events as squark itself will be heavier than the resonant
mass at HERA. Details are model dependent particularly in case of
$\tilde{t}_L$.
\end{thebibliography}
\end{document}